\documentstyle[12pt]{article}
\begin{document}

\begin{titlepage}
\begin{center}
{\bf Calculation of the Observationally Small Cosmological
Constant in the Model of Six-Dimensional Warped Brane-Bolt World}
\footnote{Talk presented at the Third International Sakharov
Conference on Physics, June 24-29, 2002, Moscow, Russia;
www.sakharov.lpi.ru}

\vspace{1cm}

Boris L. Altshuler \footnote{E-mail: altshuler@mtu-net.ru \& altshul@lpi.ru}

{\it Theoretical Physics Department, P.N. Lebedev Physical Institute,
53 Leninski Prospect, Moscow, 117991, Russia}
\end{center}

\vspace{1cm}

\begin{abstract}
In the 6-dimensional model of warped
compactification based on the euclidean AdS~Reissner-Nordstrom
metric it is possible to escape artificial 4-brane's stress-energy tensor
anistropy necessary for fulfillment of Israel junction conditions by
introducing in 3+1 spacetime of the cosmological constant which
turns out to be ${}\sim G^2$ ($G$~-~Newton's constant) and acquires
the value compatible with observations.
\end{abstract}
\end{titlepage}

\vspace{5mm}

{\bf 1. Introduction}

\vspace{5mm}

Recent experimental data (see e.g. review~\cite{Starob})
appear to confirm the accelaration rate of the universe, $H$, which
is the evidence of small positive vacuum energy
$\rho_{\it vac}\cong 9 \cdot 10^{-12}(\rm eV)^4$ and hence of the
cosmological constant $\lambda$ ($c=\hbar=1$):

\begin{equation}
\label{1}
\lambda = H^2 = 8\pi G\rho_{\it vac}\cong (10^{-33}\rm eV)^{2}.
\end{equation}

At the same time "in the standard framework of low
energy physics there appears
to be no natural explanation for vanishing or extreme smallness of the
vacuum energy", as Witten put it in~\cite{Witten}, defining the problem
as "the mistery of the cosmological constant". "The problem is so
severe, -~write authors of~\cite{Dvali}, -~that it seems reasonable to put
aside all the other cosmological issues treating them as secondary and
focus completely on the cosmological constant problem". Some scientists,
including Sakharov~\cite{Sakharov}, prefer the anthropic principle approach
to explain zero or extremely small cosmological constant in the
universe where we could develop to the level of asking questions
about $\lambda$. However attempts to find a conventional scientific
resolution of $\lambda$-problem, in particular advocated
by Witten~\cite{Witten}, seem to be more challenging and inspiring
than anthropic "explanation". Certain review of mechanisms for
generating a small current value of cosmological constant is given in
Chapter 7 of \cite{Starob}. Among them the idea of cosmologically
slow decaying "cosmological constant" (this field is called
now "quintessence") would be satisfactory if it did not result
in changing with time of fundamental constants which is strongly
restricted by observations. Also it is worthwhile to note
the Zeldovich's numerological observation~\cite{Zeld} that
gravitational correction to quantum vacuum energy of particle
of mass $m$ is $Gm^6$; this gives the appropriate value of the
cosmological vacuum energy for $m$ in between electron and
proton mass (why ???). (In~\cite{Altshul} it was shown that
Brans-Dicke theory with $\phi^{-1}$ dependence of potential of
the BD-field $\phi$ describes equal to $Gm^6$ and changing
with time vacuum energy, i.e. unifies both
ideas, giving at the same time Dirac's theory of variable
gravitational interaction unexceptable experimentally).

In this paper it is  shown that Zeldovich-type "numerology" for
cosmological constant follows from dynamics of certain higher
dimensional models with branes. We shall not escape here fine-tuning
of the brane's tension well known in the Randall-Sundrum type
models. However their generalization to the models with additional
compact dimension provides additional dynamically well grounded
condition which permits "to build a bridge" between extremely
small cosmological scale of Eq.~(\ref{1}) and Planck and
elementary particles scales.

\vspace{5mm}

{\bf 2. Description of the model}

\vspace{5mm}

Following \cite{Horowitz,Louko,Burgess} we consider
euclidean AdS~Reissner-Nordstrom solution of Einstein-Maxwell
equations in 6 dimensions derived from the action:

\begin{equation}
\label{2}
S_6=M^2\int\left\{M^{2}(R-2\Lambda)-\frac{1}{4}F_{AB}F^{AB}\right\}
\sqrt{-g_6}\,d^{6}x+S_{\it (brane)},
\end{equation}
where $M$~-~"Planck mass" in 6 dimensions supposed to be of order
or somewhere below electroweek scale; $\Lambda$~-~negative bulk
cosmological constant of same order;
$x^A=\{x^{\mu},r,\varphi\}$; $A,B=0,1\ldots5$; $\mu,\nu=0,1\ldots3$;
$\varphi$~-~compact extra dimension; $F_{AB}$~-~Maxwell field;
signature is $\{-+++++\}$. The solution is described by the
metric and vector-potential of magnetic field in extra dimensions:

\begin{equation}
\label{3}
ds^{2}=M^{2}r^{2}{\tilde g}_{\mu\nu}dx^{\mu}dx^{\nu}+\frac{dr^2}
{\Delta}+{\Delta}d{\varphi^{2}},
\end{equation}

\begin{equation}
\label{4}
{A}_{\varphi}(r)=\frac{2\sqrt{2}Mb^3}{\sqrt{3}r^3}.
\end{equation}

\begin{equation}
\label{5}
\Delta=\frac{\lambda}{M^2}-\frac{a^3}{r^3}-\frac{b^6}{r^6}+\frac{r^2}{l^2};
\end{equation}

Here $l^2=10/|\Lambda|$; $\tilde g_{\mu\nu}(x)$ is
metric of observable 4-universe with positive
curvature $\tilde R=12\lambda$; $b$ characterizes the charge of the
bulk gauge field (\ref{4}).

Further on we put equal to zero "Schwarzschild"
constant $a$ in $\Delta$~(\ref{5}):

\begin{equation}
\label{6}
a=0.
\end{equation}
This hypothesis is crucial for calculating the acceptable value
of the cosmological constant $\lambda$; in case $a\neq 0$ the
proceedure considered below comes to $\lambda\sim G$, whereas
for $a=0$ much more acceptable result $\lambda\sim G^{2}$ is
obtained. Certain justification of (\ref{6}) could be received
from "energy" considerations, \cite{Horowitz}, where general
expression \cite{Hawking1} $E\sim \int N(K-K_{0})$ was applied
to calculate the energy $E$ of AdS spacetime. Here $K$ is the trace of
extrinsic curvature of co-dimension two surface of constant
time and constant assymptotically large $r$; $K_{0}$ is the same
for the background (or reference) AdS spacetime, which in
our case is characterized by $a=b=0$ in (\ref{5}) and by the
same values of $l$, $\lambda$ and period $T_{\varphi}$ of compact
coordinate as in the solution (\ref{3})-(\ref{5}). The
energy $E$ of this solution is found to be:

\begin{equation}
\label{7}
E=(3/2)M^{8}T_{\varphi}a^{3}V_{3}.
\end{equation}
The most interesting feature of this result is perhaps the
absence of contribution to $E$ from
the "Maxwell" $b$-term of $\Delta$ (\ref{5}). On the other
hand the non-zero Schwarzschild term, $a\neq 0$ in (\ref{5}), results
in $E=\infty$ because 3-volume $V_{3}$ in (\ref{7})is infinite.\footnote{In
\cite{Horowitz}, contrary to (\ref{7}), there was received
negative value
of $E$, which, as we suppose, was a result of mistakenly taken
non-constant, dependent on $r$, period of compact coordinate of the
background spacetime (cf. (3.7) in \cite{Horowitz})}.

Thus we consider model with $a=0$ in (\ref{5}). Extra
coordinate $r$ is limited from below and from above, $r_{0}<r<R$,
where $\Delta(r_{0})=0$. The point $r=r_{0}$, which is co-dimension
two surface known as "bolt" \cite{Hawking}, is regular
provided $\varphi$ has period

\begin{equation}
\label{8}
T_{\varphi}=4{\pi}(\partial{\Delta}/\partial r|_{r=r_0})^{-1}=\frac{\pi
l^{2}}{2r_{0}}, \end{equation}
where

\begin{equation}
\label{9}
r_{0}=(b^{6}l^{2})^{1/8}.
\end{equation}
(In (\ref{8}), (\ref{9}) it was taken into account that in (\ref{5})
$\lambda/M^2\ll 1$, and only higher order terms were retained).

Positive tension 4-brane located at $r=R$ limit the space via cut and
paste proceedure; it possesses one large (${}\sim R$) compact
dimension $\varphi$. We shall not seek for artificial smallest
anisotropy of the brane's stress-energy introduced e.g.
in \cite{Burgess} to satisfy Israel junction conditions
imposed upon brane's (3+1) space-time coordinates and upon
$\varphi$-coordinate. The simplest case of brane described by the single
tension parameter is considered. Difference in $r$-dependence of scale
factors $M^{2}r^{2}$ and $\Delta$ in (\ref{3}) (this difference is rather
small for large $r$ because of AdS assymptotics of $\Delta$ in (\ref{5}))
yields well known additional condition of compatibility for stationary
brane's location (see e.g. in \cite{Visser}):

\begin{equation}
\label{10}
\left.\frac{\partial}{\partial r}\left(\frac{\Delta}{r^2}\right)
\right|_{r=R}=0.
\end{equation}
It is essential that term $r^{2}/l^{2}$ of (\ref{5}) drops out
in (\ref{10}). We follow \cite{Louko} where cosmological
constant $\lambda$ in 4 dimensions was introduced to satisfy
condition (\ref{10}).

\vspace{5mm}

{\bf 3. Effective dynamics in 4 dimensions. Weak/Planck hierarchy
and calculation of the cosmological constant}

\vspace{5mm}

Condition (\ref{10}), with account of (\ref{5}), (\ref{6}),
(\ref{9}) results in the expression for $\lambda$:

\begin{equation}
\label{11}
\lambda=\frac{4M^{2}b^{6}}{R^{6}}=\frac{4M^{2}r_{0}^{8}}{l^{2}R^{6}}.
\end{equation}
This permits to calculate $\lambda$ as a function of Newton's constant $G$.

We follow the traditional approach of evaluating dynamics in 4 dimensions,
described by the action $S_4$, by inserting ansatz (\ref{3})~-~(\ref{5}),
with account of (\ref{6}), (\ref{8}), (\ref{9}), into the
action $S_6$ (\ref{2}) and integrating out extra coordinates $r,\varphi$:

\begin{equation}
\label{12}
S_4(\tilde g_{\mu\nu}(x))=\int_{r_0}^{R}(\cdots)drd\varphi=
\int\left\{\frac{1}{16\pi G}(\tilde R-2\tilde\lambda)\right\}
\sqrt{-\tilde g}\,d^{4}x.
\end{equation}
wherefrom:

\begin{equation}
\label{13}
\frac{1}{16\pi G}=M^{6}T_{\varphi}\int_{r_0}^{R}r^{2}\,dr\cong (1/3)M^{6}
T_{\varphi}R^{3}=\frac{\pi M^{6}l^{2}R^{3}}{6r_{0}}.
\end{equation}

Thus $G^{-1}$ being ${}\sim R^{3}$ is "diluted" in extra space.
Week/Planck hierarchy is obtained if matter fields are trapped
upon the additional "week scale" brane placed at $r=r_{0}$, or
if fields describing the masses of matter fields are, contrary
to the Planck scale, "concentrated" near $r=r_{0}$. The example
of such a "mass field" concentrated near lower values of $r$ may
be vector-potential (\ref{4}). It may provide mass to the charged
fields interacting with $A_{B}$, or, in case $A_{\varphi}$ is a
component of some non-abelian gauge field, presence of $A_{\varphi}^2$ in
quartic terms of the Yang-Mills action will result in breakdown of gauge
symmetry in 4 dimensions; then some components of the gauge field will
acquire masses. These possibilities deserve further investigation.

Cosmological constant $\tilde\lambda$ in $S_4$ (\ref{12}) is given by the
bulk value of action $S_6$ (\ref{2}) which, according to the well known
consistency condition, coincide on the solution of dynamical equations
with input cosmological constant $\lambda$ of metric $\tilde g_{\mu\nu}$
in (\ref{3}), (\ref{5}) (the proof see in \cite{Forste}):

\begin{equation}
\label{14}
\tilde{\lambda}=\lambda.
\end{equation}

To receive final expression for cosmological constant in 4 dimensions
we substitute $R$ by $G$ from (\ref{13}) into the
expression (\ref{11}) for $\lambda$, it is convenient
also to express $\lambda$ not by "charge" $b$ but by more
"geometrical" constant - minimal value $r_{0}$ of the radial
coordinate, connected with $b$ via (\ref{9}):

\begin{equation}
\label{15}
\lambda=(16\pi^{2}/3)^{2}G^{2}M^{14}l^{2}r_{0}^{6}.
\end{equation}

If fundamental scale $M$ and "length" $l$ which determines the
bulk cosmological constant in the action (\ref{2}) ($\Lambda=-10/l^{2}$)
are supposed to be of the electroweek scale $M=l^{-1}=1 \rm TeV$,
then observable value (\ref{1}) of $\lambda$ is
obtained from (\ref{15}) for $r_{0}=10^{-22} \rm cm$.

If on the other hand all three low energy scales in (\ref{15}) are equal:

\begin{equation}
\label{16}
M=l^{-1}=r_{0}^{-1}\equiv m,
\end{equation}
then Zeldovich formula \cite{Zeld} for vacuum
energy $\rho_{\it vac}$ follows from (\ref{15}), (\ref{16}):

\begin{equation}
\label{17}
\rho_{\it vac}=\lambda/8\pi G=\frac{32\pi^{3}}{9}Gm^{6}.
\end{equation}
This gives observable value of vacuum energy
for $m\cong 10^{7} \rm eV$. But if we take $r_{0}=l$ and $Ml=10^{-3}$
then observable value of $\lambda$ (\ref{1}) is obtained for the values
of parameters which are rather close to the electroweek
scale: $r_{0}^{-1}=l^{-1}=10^{14}\rm eV$, $M=10^{11}\rm eV$. The
game with numbers may be continued. It does not make much sense
however before the physical meaning of constants of the model and
their possible connections with elementary particles masses are understood.

\vspace{5mm}
\newpage

{\bf 4. Comments}

\vspace{5mm}

Contrary to the conventional situation in warped models
with extra dimensions where $\lambda$-term of the 4-dimensional
universe is an arbitrary constant of solutions of dynamical
equations, in the model considered here this constant is not
an arbitrary one but, because of sophisticated Israel junction
conditions, is calculable through parameters ($M, \Lambda$) of the
theory and constants ($R, r_{0}$) of the solution; the same is
true for Newton's constant (see final expressions in (\ref{11}),
(\ref{13})). Thus any physical mechanism which defines (stabilizes)
the boundaries $R, r_{0}$ of the extra space will simultaneously
determine the mass hierarchy and value of the cosmological constant.

Essential drawback of the model is fine-tuning of the
brane's tension, which is proportional to $\sqrt{\Delta(R)}/R$
\cite{Louko}, and hence includes corrections depending on extremely
small value of $\lambda/M^{2}$. It would be interesting to study the
"self-tuning" theory with additional scalar field dependence in the
bulk and brane actions in (\ref{2}).

\vspace{5mm}

{\bf Acknowledgement}
This work was supported in part by the Russian Foundation for
Basic Research (project no. 00-15-96566).

\end{document}